\documentclass[%
 reprint,
 amsmath,amssymb,
 aps,
]{revtex4-2}

\usepackage{graphicx}
\usepackage{dcolumn}
\usepackage{bm}
\usepackage{color,soul}
\usepackage{amsmath}
\usepackage[utf8]{inputenc}
\usepackage[english]{babel}
\usepackage{setspace}
\usepackage{color,soul}
\usepackage{xcolor}

\usepackage{chngcntr}
\counterwithout{figure}{section}
\counterwithout{equation}{section}

\begin{document}


\title{Reconfigurable multifunctional metasurfaces employing hybrid phase-change plasmonic architecture}

\author{Sajjad Abdollahramezani}
\author{Hossein Taghinejad}
\author{Tianren Fan}
\author{Mahmood Reza Marzban}
\author{Ali A. Eftekhar}
\author{Ali Adibi}

\email{Email: ali.adibi@ece.gatech.edu}

\affiliation{School of Electrical and Computer Engineering, Georgia Institute of Technology, 778 Atlantic Drive NW, Atlanta, Georgia 30332-0250, US}

\date{\today}

\begin{abstract}

We present a hybrid device platform for creating an electrically reconfigurable metasurface formed by the integration of plasmonic nanostructures with phase-change material germanium antimony telluride (GST). By changing the phase of GST from amorphous to crystalline through Joule heating, a large range of responses from the metasurface can be achieved. Furthermore, by using the intermediate phases of GST, the metasurface can interact with the incident light in both over-coupling and under-coupling regimes, leading to an inherently broadband response. Through a detailed investigation of the nature of the fundamental modes, we demonstrate that changing the crystalline phase of the GST at the pixel-level enables an effective control over the key properties (i.e., amplitude, phase, and polarization) of incident light. This leads to the realization of a broadband electrically tunable multi-functional metadevice enabling beam switching, focusing, steering, and polarization conversion. Such a hybrid structure offers a high-speed, broadband, and non-volatile reconfigurable paradigm for electrically programmable optical devices such as switches, holograms, and polarimeters.

\end{abstract}


\maketitle


\section{Introduction}

In recent years, there has been an increasing interest in metasurfaces, the planar array of optically-thin patterned scatterers distinguished through their induced electric and magnetic surface currents. Metasurfaces are capable of abruptly changing the local state of light thanks to their engineered subwavelength unite cells, namely meta-atoms \cite{yu2011light,sun2012high}.
Reasonable dissipation loss, ease of fabrication, amenability for implementation of tunable devices, and compatibility with the widespread CMOS fabrication technology make metasurfaces an appealing flat platform for tailoring the spatial, spectral, and temporal properties of optical wavefronts \cite{kuznetsov2016optically,hemmatyar2019full,arbabi2016multiwavelength,abdollahramezani2015beam,abdollahramezani2020meta}.

To date, extensive research has been conducted on static metasurfaces in which the geometry and compositions of subwavelength features of a metasurface are elaborately designed for a specific functionality that remain unchanged during the operation \cite{dorrah2022tunable,abdollahramezani2020meta}. However, to efficaciously extend the unique capabilities of metasurfaces to state-of-the-art applications such as adaptive optics, imaging, and microscopy, the development of reconfigurable metasurfaces is becoming indispensable. 

So far, several approaches have been proposed in which external stimuli including chemical reactions, heat, elastic forces, magnetic fields, optical pumping, and electrostatic fields are applied to actively adjust the optical properties of metasurfaces \cite{ferrera2017dynamic,makarov2017light}. Among these, dynamic electro-optical metasurfaces incorporating active functional materials such as transition metal dichalcogenides, liquid crystals, graphene, phase-transition materials, highly doped semiconductors, and transparent conducting oxides show superiority in terms of the tuning range, response time, energy consumption, robustness, and pixel-by-pixel reconfiguration. 
Nevertheless, fabrication complexities in the realization of defectless uniform transition metal dichalcogenides \cite{sun2016optical}, surface anchoring in optically thick liquid crystal-based metasurfaces  \cite{franklin2015polarization}, high inherent loss of graphene in near-infrared (near-IR) and visible (VIS) wavelengths \cite{yao2013wide,abdollahramezani2015beam}, the restricted tuning range and volatility of VO$_{2}$ \cite{wang2015switchable}, and the low accessible index contrast of doped semiconductor-based metasurfaces \cite{iyer2015reconfigurable} introduce serious impediments in the realization of reconfigurable metasurfaces with practically desired tunning range and/or speed.   
Despite the advantages of transparent conducting oxides, specifically indium tin oxide (ITO), including ultrafast modulation speed ($\sim$ fs \cite{taghinejad2018hot}) and tunable electro-optical properties through the pre-/post-deposition processes \cite{kafaie2018dual}, the large unity-order index variation is restricted to the near-IR wavelengths and achieved at the expense of dramatic dissipative losses within the exceptional epsilon-near-zero operation region. Furthermore, their ultra-thin ($\sim1$  nm) inhomogeneous active layer strictly degrades the reflection efficiency, limiting their functionalities to the near-IR regime.
Table S1 in the Supporting Information provides a comprehensive comparison between the recently reported reconfigurable metasurfaces with different control mechanisms.

Due to the subwavelength pixel size of metasurfaces (that limits the light-matter interaction length and time) and the moderate quality-factor (Q) of their constituent resonators (that limits the field enhancement), dramatic change in the complex refractive index of the metasurface constituents is necessary to form practically viable reconfigurable metasurfaces. For this purpose, the development of alternative functional materials with notable refractive index change upon excitation with external stimuli and the capability to form compact and scalable platforms for high-speed and power-efficient architectures are highly desirable. 

Phase-change materials (PCMs), as the spotlight of rewritable optical data storage disks and electronic memories, offer a striking portfolio of properties for realization of reconfigurable metasurfaces. Amongst them, semiconducting germanium antimony tellurium (Ge$_{2}$Sb$_{2}$Te$_{5}$ or shortly GST), has lately garnered widespread interest in the emerging metaphotonic applications \cite{wuttig2017phase}. The refractive index of GST can be changed by a large amount \cite{abdollahramezani2020tunable} ($> 2.3$ in near-IR) through changing its phase from amorphous to crystalline. In addition, GST is of great interest due to its subwavelength scalability (down to the nanometer size), ultrafast switching speed (picosecond or less), high switching robustness (potentially up to 10$^{15}$ cycles), low power consumptive non-volatility, high thermal stability, and compatibility with the CMOS fabrication technology \cite{wuttig2017phase,abdollahramezani2020tunable}. More importantly, the refractive index of GST can be selectively controlled within the intermediate phases between amorphous and crystalline states rendering remarkably different optical and electrical characteristics. 

So far, most studies of PCM-based metasurfaces have been based on tailoring the amplitude response of incident light \cite{wuttig2017phase,abdollahramezani2020tunable}, restricting the possible functionalities to free-space optical switching \cite{gholipour2013all}, perfect absorption \cite{chen2015tunable}, imaging \cite{hemmatyar2021enhanced}, thermal emission \cite{qu2017dynamic}, and harmonic generation \cite{zhu2021dynamically,ZhuAbdollahramezani2022}. Very recently, a growing attention has been drawn to the phase-front manipulation with applications in beam steering and beam shaping \cite{zhang2021electrically,abdollahramezani2021dynamic,abdollahramezani2021electrically,wang2021electrical}. The major drawbacks of such conventional PCM-based metasurfaces are three fold. First, each designed structure is only capable of manipulating a specific property of the optical wavefront (i.e., amplitude, phase, polarization, or spectrum), thus limiting the beam reconfigurability potential. Second, the post-fabrication modulation mostly relies on the focused optical pulses or bulky thermal heaters, as common time-controlled external stimuli, thus hindering the feasibility of device integration. Finally, the phase transition of the PCM inclusion is mostly occurred over the whole area of the metasurface with no pixel-level addressability. 

In this paper, we demonstrate a reconfigurable hybrid metasurface by combining the unique features of plasmonic nanoantenna arrays with unmatched tuning properties of a PCM. We leverage the multi-level non-volatile partial crystallization of GST realized through Joule heating in individually electrically-controlled meta-atoms. By modulating the optical state of the incident light over the subwavelength scale, manifold functionalities are demonstrated with an optimized reconfigurable metasurface.  
Finally, we perform transient electrothermal simulations to monitor the effect of temporal temperature change on the fast conversion mechanism of GST upon applying an electric stimulus. 


\section{Structure design, functional materials, and underlying physics}

\begin{figure} [b]
	\centering
	\includegraphics[trim=13cm 3Cm 10.5cm 0cm,width=7.3cm,clip]{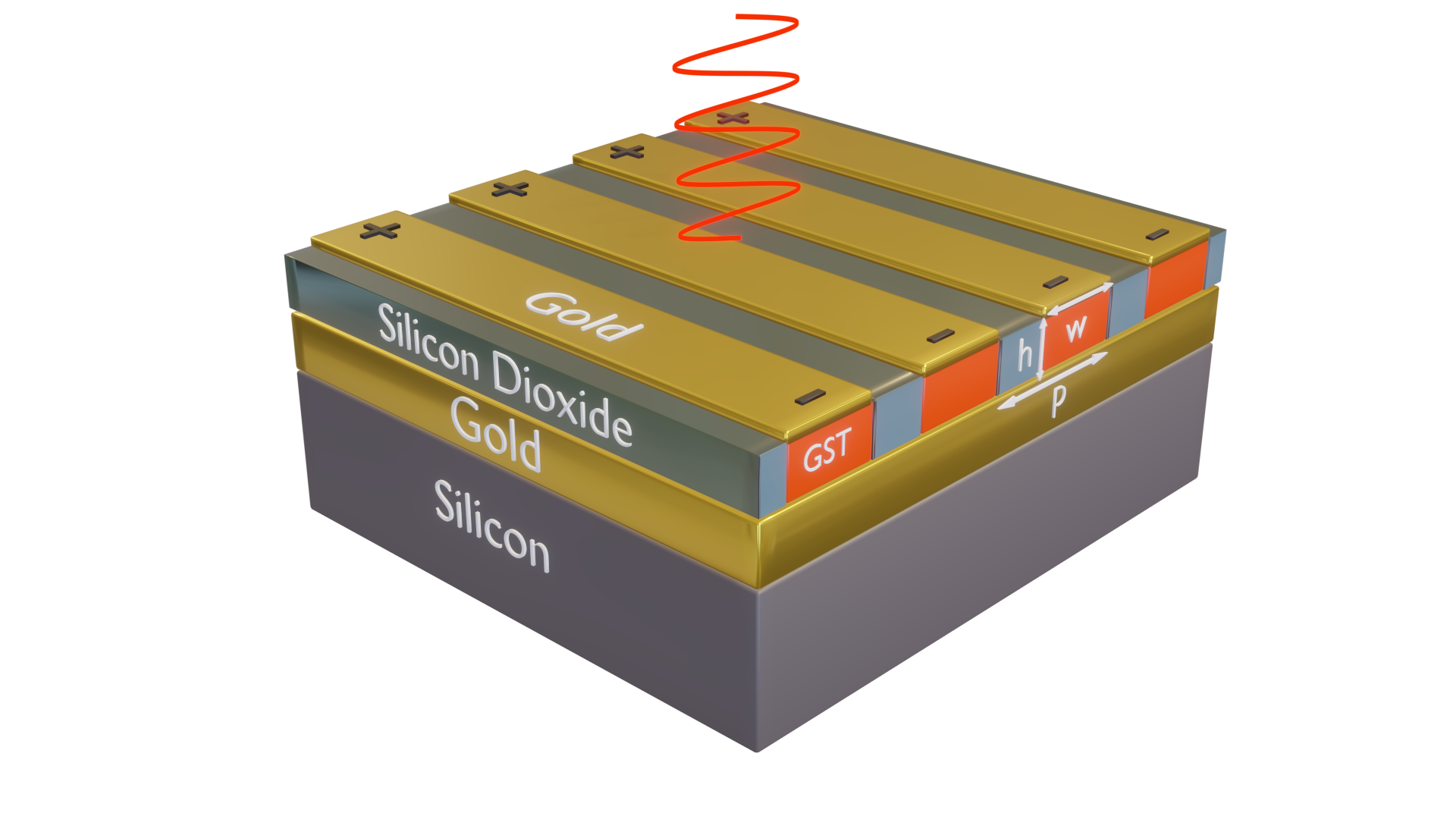}\\
	\caption
	{Three-dimensional illustration of the proposed electrically reconfigurable metasurface. The metasurface structure consists of patterned Au nanoribbons on the GST nanostripes, which are separated from each other by SiO$_{2}$ spacers, deposited on an Au back reflector. The entire structure resides on a Si substrate with 500$~\mathrm{\mu}$m thickness. A voltage signal can be applied between two end-faces of the Au nanoribbon to partially convert the corresponding GST nanostripe in the desired nanoantenna from the amorphous to the crystalline states (or any state in between). The periodicity of the meta-atom is p = 550 nm while the width and thickness of each Au nanoribbon are w = 340 nm and 30 nm, respectively. The thickness of GST nanostripes is h = 180 nm. Throughout this paper, all geometrical dimensions (p, w, and h) remain unchanged.
	}
	\label{fig.3D perspective}
\end{figure}

Figure \ref{fig.3D perspective} depicts the structure of the proposed electrically tunable hybrid metasurface. The structure is comprised of a one-dimensional (1D) array of periodic gold (Au)/GST nanoantennae, which are separated by silicon dioxide (SiO$_{2}$) spacers, deposited on an optically thick Au back-reflector. Here, the pivotal structural parameters, i.e., the height (h) and width (w) of the GST nanostripes, the periodicity (p) of the array, and the thickness (t) of Au nanoribbons are optimized to realize a multifunctional metasurface. Reconfiguration is achieved by the dynamic control over the optical properties of the structure granted by selectively adjusting the crystalline phase of GST nanostripes. To do so, Au nanoribbons can be connected individually or as a cluster (associated with the desired functionalities) to separate in-plane electrodes. Then, electrical gating along the top nanoribbon is realized to trigger the thermal annealing and melting processes necessary for implementation of repeatable multi-level crystallized states of the GST nanostripes (see Section 4).

The ellipsometrically measured complex permittivity of GST in amorphous and crystalline phases are shown in Figure S1 \cite{shportko2008resonant}. While the larger dielectric constant of the crystalline state stems from highly polarizable delocalized resonant bonding, the lower dielectric constant of the amorphous phase is mostly ascribed to covalent bonds \cite{wong2010phase,wuttig2017phase}. Transition between these two above-mentioned extreme phases is practically accomplished by an external stimulus, with specified power for a pre-defined time duration, to generate localized Joule heating in the GST nanostripe (see Section 4). Accordingly, in the intermediate states, GST can be viewed as an arbitrary spread of amorphous and crystalline elements \cite{raoux2009phase,loke2012breaking}. Among several effective-medium theories proposed for the description of the dielectric function of heterogeneous media, we utilize the well-known Lorentz-Lorenz relation \cite{chu2016active} to approximate the effective permittivity ($\epsilon_{eff}(\lambda)$) of partially crystallized GST as follows:
\begin{equation} 
\frac{\epsilon_{eff}(\lambda)-1}{\epsilon_{eff}(\lambda)+2}={L_{c}}\times\frac{\epsilon_{c}(\lambda)-1}{\epsilon_{c}(\lambda)+2}+({L_{c}}-1)\times\frac{\epsilon_{a}(\lambda)-1}{\epsilon_{a}(\lambda)+2}
\label{Lorentz-Lorenz}
\end{equation}
where for a specific wavelength $(\lambda)$, $\epsilon_{c}(\lambda)$ and $\epsilon_{a}(\lambda)$ are the permittivities of crystalline and amorphous GST, respectively, and ${L_{c}}$, ranging from 0 (amorphous) to 1 (fully crystalline), is the crystallization fraction of GST. 

\begin{figure}[t]
	\centering
	\begin{tabular}{@{}cccc@{}}
		\includegraphics[trim=0cm 11.5cm 11cm 0cm,width=8.3cm,clip]{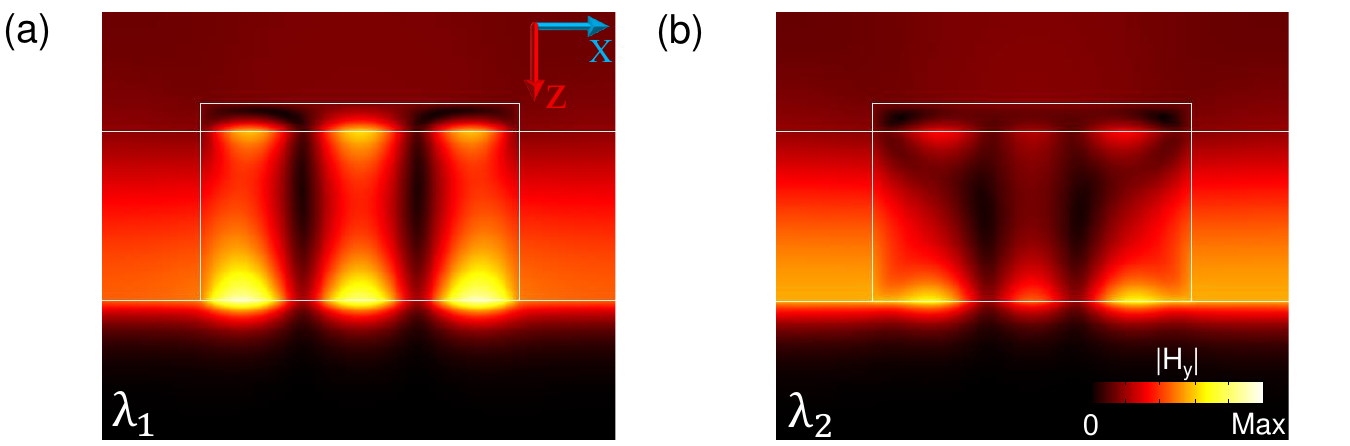}\\
		\includegraphics[trim=0cm 6cm 0cm 0cm,width=8.3cm,clip]{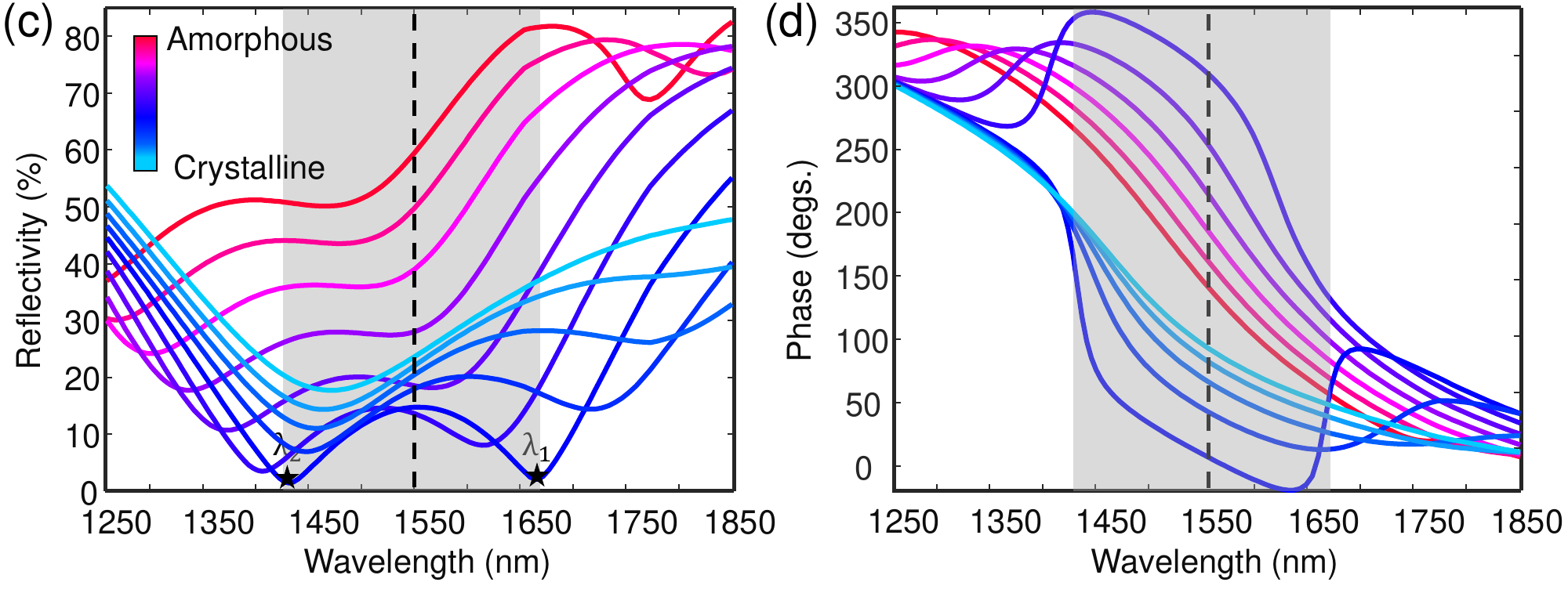}\\
	\end{tabular}
	\caption{Reflection response of the reconfigurable metasurface in Figure \ref{fig.3D perspective}. Magnetic field intensity profiles of the (a) SR-SPP, and (b) PR-SPP modes excited within the cross-sectional view of the meta-atom (in Figure \ref{fig.3D perspective}(b)) with $L_{c}^{60}$ (corresponding to $60\%$ crystallization fraction). (c) Reflection amplitude spectra of the metasurface in Figure \ref{fig.3D perspective} for several crystallization fractions. Upon partial to ultimate conversion of the phase of GST nanostripe from amorphous to the crystalline, the SR-SPP ($\lambda_{1}$) and PR-SPP ($\lambda_{2}$) resonance wavelengths can be gradually tuned resulting in different reflectivities. (d) Corresponding reflection phase spectra of the metasurface. The wavelength region in which the fundamental modes of the structure interact with each other is highlighted with the gray shaded region.}
	\label{fig.reflection and profiles}
\end{figure}

To analyze the proposed hybrid plasmonic-GST metasurface in Figure \ref{fig.3D perspective}, numerical simulations were conducted by the finite-element method (FEM) and the finite-integral technique (FIT) with realistic material parameters (see Section S1 in the Supporting Information). Figure \ref{fig.reflection and profiles} depicts the scattering properties of the proposed meta-atom with p = 550, w = 340 nm, t = 30 nm, and h = 180 nm assuming that the GST nanostripe crystallization level (${L_{c}}$) varies from 0 to 1, when the structure is excited by a normal transverse-magnetic (TM) polarized (i.e., magnetic field is perpendicular to the grating direction) plane wave of light.

The metasurface in Figure \ref{fig.3D perspective} is designed to have two modes whose profiles are shown in Figures \ref{fig.reflection and profiles}(a) and \ref{fig.reflection and profiles}(b) for a specific GST crystallization factor of $L_{c}^{60}$ (i.e., $60\%$ crystallization). The long-wavelength mode corresponds to the local resonance of the lateral plasmonic Fabry-Pero$\mathrm{\acute{t}}$ (F-P) mode of the Au/GST nanoantenna, namely the short-range surface plasmon polariton (SR-SPP) working in the over-coupling regime. On the other hand, the short-wavelength mode is the resonance of the 1D grating structure of the metasurface at propagating wavevector (k = 0), called propagative surface plasmon polariton (PR-SPP), operating in the under-coupling condition.
The interaction of the two fundamental modes of the metasurface highly depends on the crystallization state of the GST nanostripe leading to a large dynamic range in the reflection spectrum. This is shown in Figure \ref{fig.reflection and profiles}(c) and \ref{fig.reflection and profiles}(d) for the amplitude and phase of the reflected wave from the metasurface, respectively.

\begin{figure} [t]
	\centering
	\includegraphics[trim=12cm 4cm 0cm 0cm,width=7.3cm,clip]{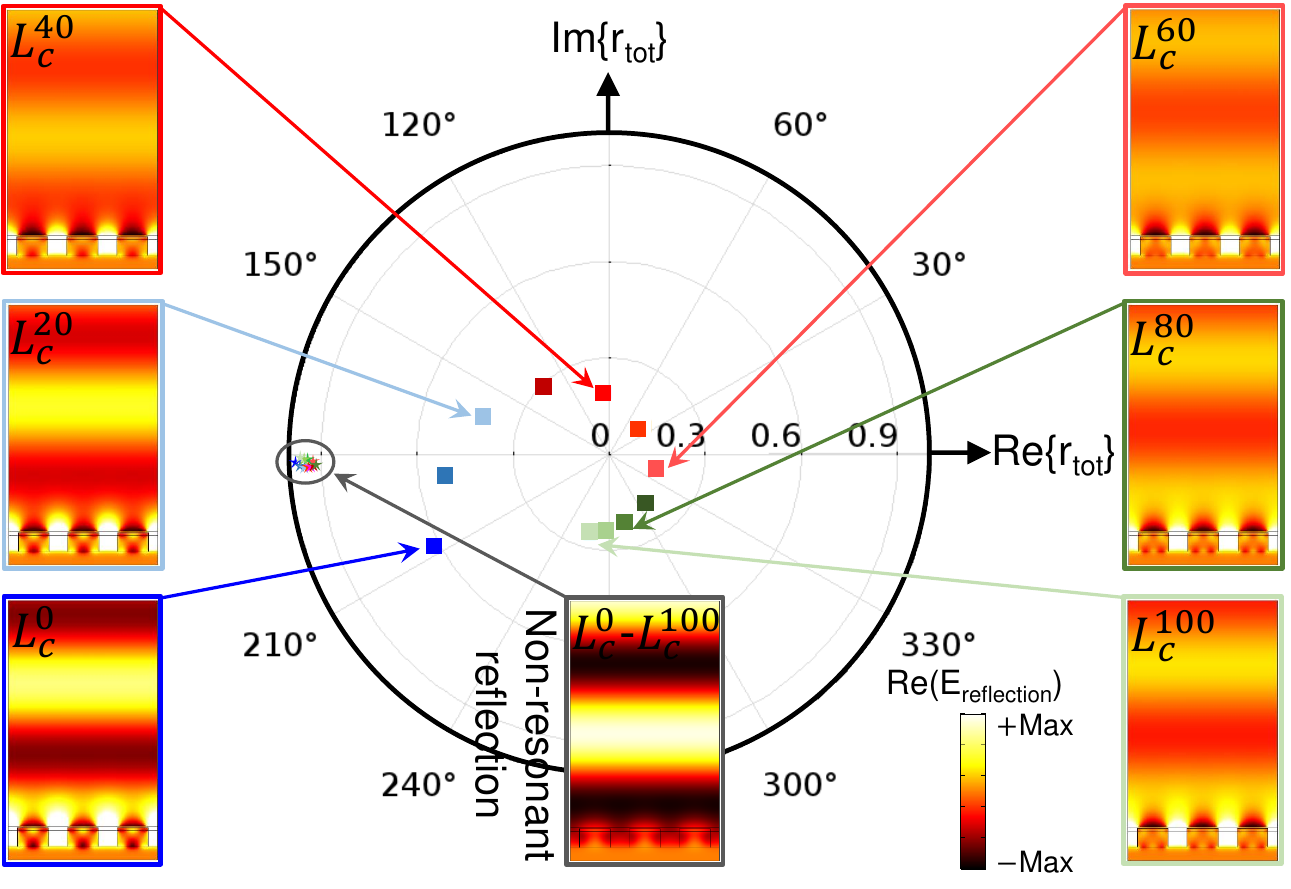}\\
	\caption
	{Polar plot representation for the trajectory of the total complex reflection coefficient (r$_{\textrm{tot},L_{c}}$) with increasing the GST crystallization level. The condition of metasurface moves form the over-coupling for near-amorphous phase (bluish dots) towards the under-coupling state in the intermediate phases (reddish dots), and it again enters the over-coupling regime when the GST phase becomes close to crystalline (greenish dots). The non-resonant reflection responses accommodated in the center gray box confirm the negligible variation of both amplitude and phase for all crystallization levels (depicted with stars in the gray oval). To compare the phase shift between different GST crystallization levels, the corresponding electric field profiles of resonant and non-resonant reflected light are illustrated. $L_{c}^{\textmd{t}}$ represents the crystallization fraction of t$\%$ for the GST nanostripe of each resonator (i.e., $L_{c}^{0}$ and $L_{c}^{100}$ correspond to the amorphous and crystalline phases, respectively).
	}
	\label{fig.polar plot}
\end{figure}

The SR-SPP mode, a tightly-confined slow propagative electromagnetic wave coupled to the collective electron oscillations of the Au nanoribbon, is excited at the interface of nanoribbon and GST and propagates back and forth between the two Au end-faces \cite{barnard2008spectral}. When the SR-SPP mode experiences the truncation, it partially reflects and loses a fraction of its energy to the free-space scattered field. An Au nanoribbon with an appropriately chosen width can be treated as a F-P resonator generating a constructive interference between forward and backward traveling SR-SPP modes leading to a highly confined, higher mode-index resonant field adjacent to the nanoribbon. The resonance condition is given by \cite{schuller2010plasmonics}:
\begin{equation} \textmd{w}=\frac{m\pi-\phi(\epsilon_{\textrm{eff}})}{2\pi}\frac{\lambda_{1}}{n_{\textrm{SR-SPP}}(\epsilon_{\textrm{eff}})}
\label{resonance condition}
\end{equation}
where ${n_{\textrm{SR-SPP}}(\epsilon_{\textrm{eff}})}$ is the index of the SR-SPP mode freely propagating at the interface of an Au film with the same thickness as the nanoribbon, $m$ is a positive integer indexing the resonance order, and $\phi(\epsilon_{\textrm{eff}})$ is the phase change due to the reflection at the end-faces of the F-P structure. It should be underlined that both $\phi(\epsilon_{\textrm{eff}})$ and ${n_{\textrm{SR-SPP}}(\epsilon_{\textrm{eff}})}$ in a general F-P resonator are functions of the dielectric permittivities of Au and GST. Although eq. \ref{resonance condition} provides a physical insight for explaining the resonance condition, calculating the exact wavelength of the resonant mode requires accurate calculation of $\phi(\epsilon_{\textrm{eff}})$, which is complicated especially when the surrounding material is dispersive and lossy. It is worth mentioning that decreasing the gap size between the Au nanoribbon and back reflector pushes the SR-SPP to confine more tightly to the Au surface, leading to the domination of the absorption. Meanwhile, due to the large refractive indices of GST in different phases, by decreasing the gap size only a small portion of the incident light couples to the gap surface plasmons. As a result, the electromagnetic field enhancement is so weak that by modifying the crystallization fraction, a negligible change in the reflection spectrum is expected. To overcome this issue, the design of GST nanostripes with an optimized height is indispensable. Figure \ref{fig.reflection and profiles}(a) demonstrates the coupling between third-order (m = 3) SR-SPPs of the Au nanoribbon and the enhanced SR-SPPs-like mode excited at the surface of the back reflector, which results in the enhanced light-matter interaction in the GST nanostripe. Here, the absorption mechanism can be explained by the Ohmic losses in the Au nanoribbon and the back reflector as well as the energy absorbed in the GST nanostripe upon multiple reflections of the standing wave between subsequent SiO$_{2}$ spacers.  

The origin of the second mode and thus the absorption dip at $\lambda_{2}$ can be predicted according to the grating equation
\begin{equation} 
{\frac{2\pi}{c\lambda_{2}}\textrm{sin}(\theta)\pm{q\frac{2\pi}{p}}=\pm{\beta}}
\label{grating condition}
\end{equation}
where $\theta$ is the incidence angle (with respect to the normal line), $q$ is the grating diffraction order, and $\beta$ is the wavenumber of the propagating mode along the interface of the Au back reflector. A simple inspection of the symmetric mode profile in Figure \ref{fig.reflection and profiles}(b) confirms that the PR-SPP mode roots in the two-way propagation of the excited SPPs, which tunnel into the GST nanostripe. It is worth mentioning that the diffraction from the sharp edges of the individual Au nanoribbons can provide the extra momentum required to generate the PR-SPP under the normal excitation (i.e., $\theta=0$) \cite{pors2013efficient,doi:10.1021/acsphotonics.7b00959}. Moreover, Figures \ref{fig.reflection and profiles}(a) and \ref{fig.reflection and profiles}(b) verify that PR-SPPs do not have similar enhancement and localization degree to those of SR-SPPs primarily because Au nanoribbons only perturb the incoming light with slightly confined field in GST nanostripes. In this case, the absorption is mostly caused by the Ohmic loss of the metallic back reflector. In the Supporting Information, we justify the presence of the aforementioned modes by further investigating the evolution of resonances upon modification of the geometrical parameters of the metasurface and the angle of incidence.

Considering the characteristics of two fundamental modes of the metasurface in Figure \ref{fig.3D perspective}, we carry out full-wave simulations to investigate the evolution of the complex reflection coefficient representing the optical behavior of the metasurface. For the sake of brevity, we assume the metasurface as a closed optical system that limits the energy exchange through a channel. Basically, the total reflection coefficient (r$_{\textrm{tot},L_{c}}$) is governed by the non-resonant and resonant responses of the structure represented by \cite{park2016dynamic} 
\begin{equation} 
{r_{\textrm{tot},L_{c}}=r_{\textmd{non-res},L_{c}}\textmd{exp}(i\theta_{\textmd{non-res},L_{c}})+r_{\textmd{res},L_{c}}\textmd{exp}(i\theta_{\textmd{res},L_{c}})}
\label{non-resonant and resonant responses}
\end{equation}
in which the subscript $L_{c}$ denotes the level of GST crystallization, and ``$\textmd{non-res}$'' and ``$\textmd{res}$'' subscripts stand for non-resonant and resonant, respectively. Figure \ref{fig.polar plot} shows the polar plot of for the trajectory of the complex reflection coefficient of the metasurface in Figure \ref{fig.3D perspective} for different crystallization states of GST nanostripes. According to Figure \ref{fig.polar plot}, the non-resonant term is almost intact, i.e., $r_{\textmd{non-res},L_{c}}\textmd{exp}(i\theta_{\textmd{non-res},L_{c}})\approx-1$, for all levels of crystallization. We attribute this term to the reflection from an effective metallic plate, which is reasonable due to the high impedance characteristic of Au in near-IR. The field profile of the non-resonant contribution justifies this assumption as the field strength above the Au nanoribbons is much larger than that below the nanoribbons. Thus, the non-resonant reflection phase remains constant (i.e., $180^{\circ}$), and its field profile in Figure \ref{fig.polar plot} can serve as a reference for the remaining discussion. On the other hand, changing the crystalline state of the GST nanostripes significantly impacts the resonant response and thus, the resonant phase ($\theta_{\textmd{res},L_{c}}$ in eq. \ref{non-resonant and resonant responses}). Figure \ref{fig.polar plot} demonstrates that upon conversion of GST from the amorphous to the crystalline state, the reflection phase evolves clockwise from 208$^{\circ}$ to -107$^{\circ}$ while the reflection amplitude (or equivalently scattering cross section) first decreases (up to about $L_{c}^{60}$) and then increases on the GST phase becomes closer to crystalline. 
Comparing the electric field profiles with the reference profile reveals that the system is mostly in an over-coupling state, in which the resonant response dominates the non-resonant one near the amorphous state (bluish dots). The resonant response moves toward the under-coupling regime (reddish dots), where the non-resonant response overcomes the resonant one. The response finally comes back to the over-coupling regime (greenish dots) when approaching the crystalline phase. Such fundamental variation of the coupling regime around the operational wavelength, i.e., 1550 nm, due to GST phase change results in a broadband response (see Figure \ref{fig.reflection and profiles}(d)), which is a unique feature of the metasurface in Figure \ref{fig.3D perspective}.

\begin{figure} [t]
	\centering
	\includegraphics[trim=12.5cm 0cm 0cm 0cm,width=7.3cm,clip]{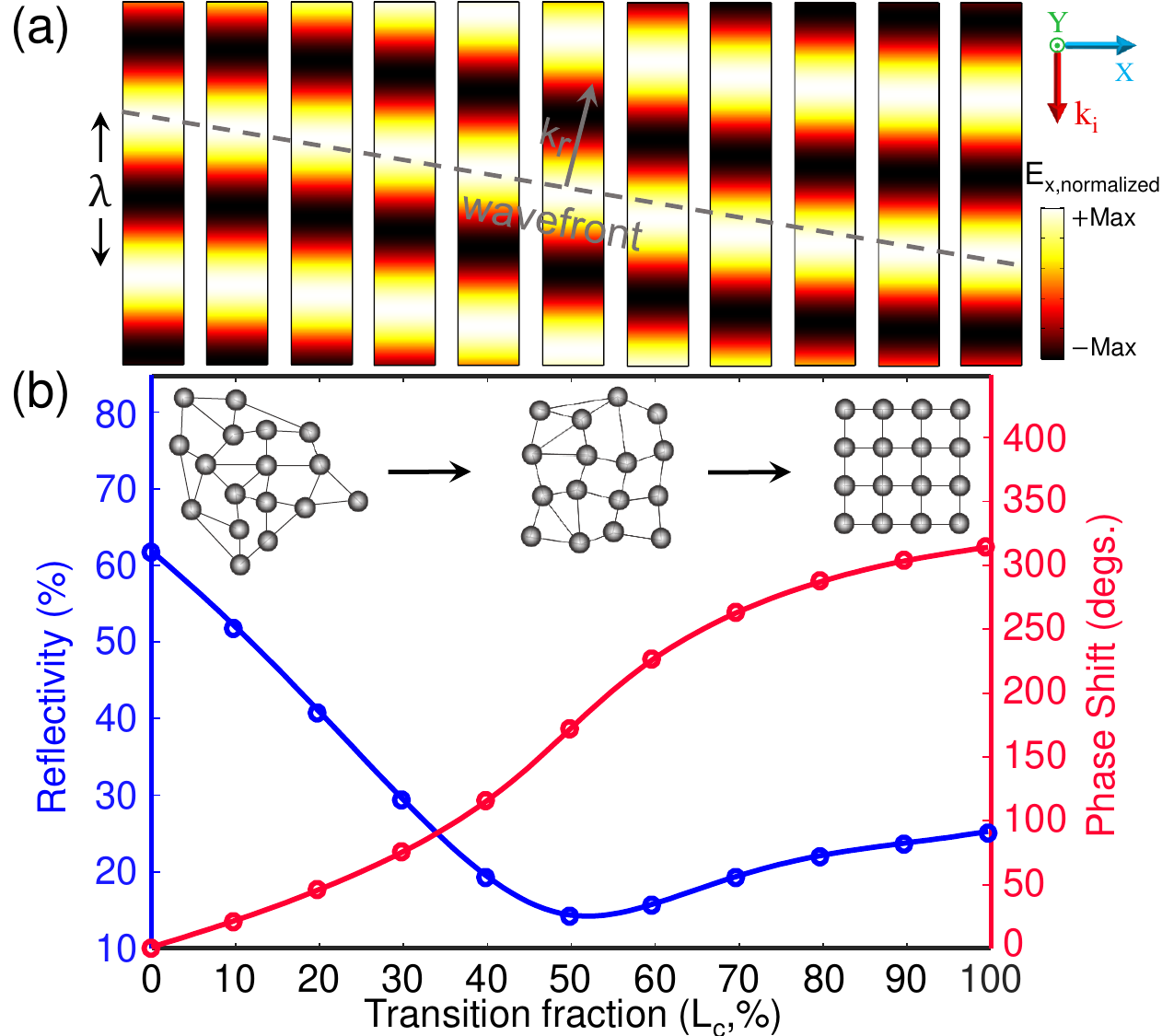}\\
	\caption
	{Amplitude and phase evolution of the hybrid meta-atom in Figure \ref{fig.3D perspective} upon conversion from the amorphous to the crystalline state. (a) Normalized reflected $E_{x}$ field patterns from the metasurface in Figure \ref{fig.3D perspective} for eleven GST crystallization states (i.e., amorphous (leftmost), fully crystalline (rightmost), and 9 intermediate states separated by 10$\%$ crystallization from each other). Each pattern is calculated by an independent full-wave numerical simulation under the illumination of a normally x-polarized plane wave with $\lambda=1550$ nm (see Figure \ref{fig.3D perspective}). The gray dashed line defining the wavefront verifies the achievable total phase shift of $315^{\circ}$. (b) Evolution of reflection amplitude (left axis) and phase shift (right axis) as a function of the GST crystallization fraction. Inset: conceptual schematic depicts the gradual non-volatile phase transition of GST from amorphous (left) to fully crystalline (right). All design parameters are the same as those in the caption of Figure \ref{fig.3D perspective}.
	}
	\label{fig.eleven crystallization levels}
\end{figure}

\begin{figure*} [t]
	\centering
	\includegraphics[trim=7.5cm 0cm 0cm 0cm,width=11cm,clip]{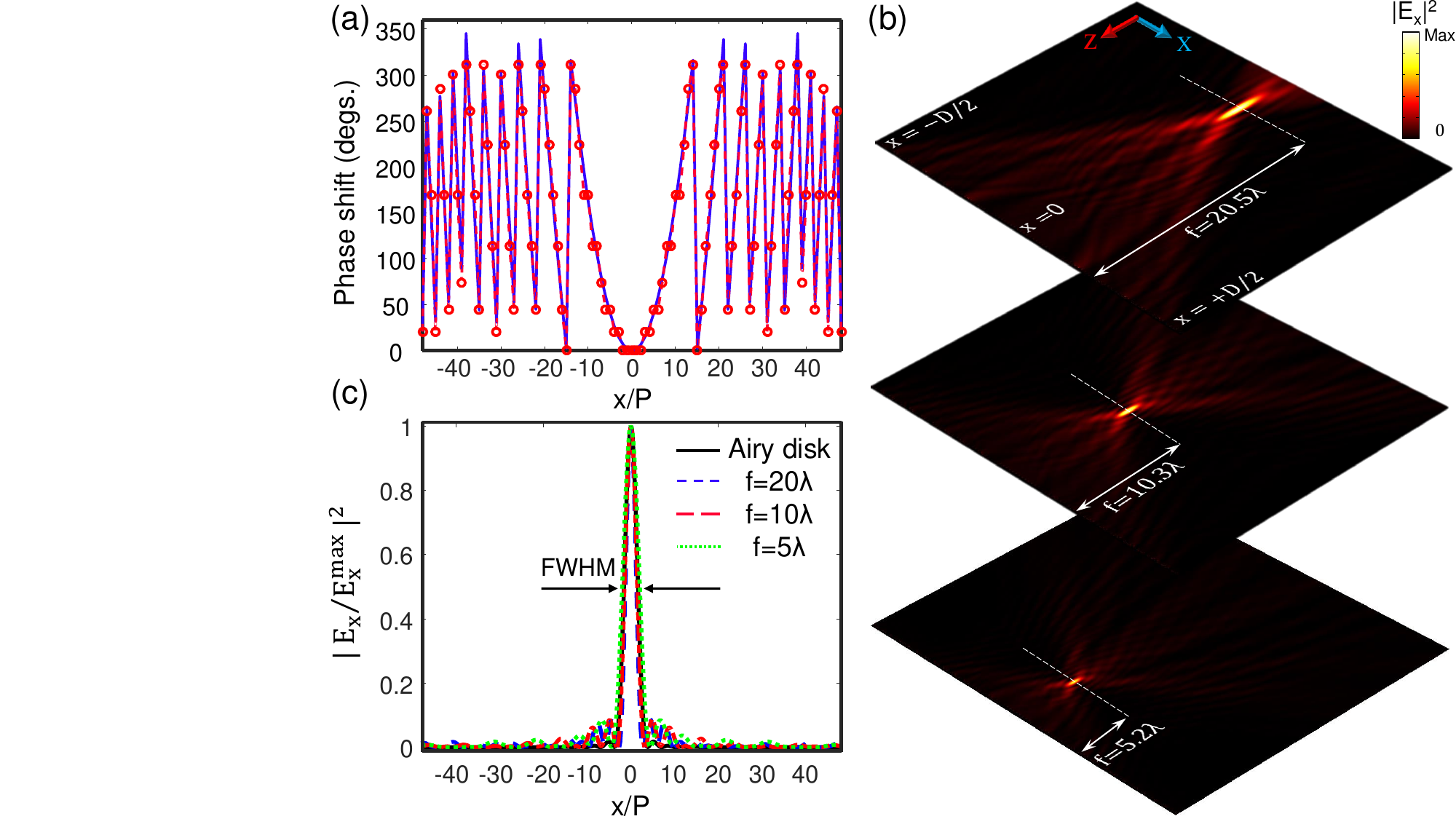}\\
	\caption
	{Performance of a reconfigurable metalens with diverse focal lengths. (a) The desired (solid blue line) and realized (red circles) phase discontinuities to implement a metalens with the focal length of 20$\lambda$. (b) Calculated intensity profiles of the reflected beam by the tunable metalens in the transverse cross section for nominal focal lengths of (top)   f = 20$\lambda$, (middle) f = 10$\lambda$, and (bottom) f = 5$\lambda$.  (c) Comparison of the Airy disk intensity (solid black line) with the squared normalized amplitude of the electric field calculated at the distances of  f = 20$\lambda$, f = 10$\lambda$, and f = 5$\lambda$ (dashed lines). The lateral dimension of lenses is D = 100p, and the dashed horizontal lines in (b) indicate the focal plane. The wavelength of the linearly-polarized incident light is $\lambda=1550$ nm.
	}
	\label{fig.focusing}
\end{figure*}

\section{Implementation of a reconfigurable multifunctional metadevice}

Figure \ref{fig.eleven crystallization levels}(a) shows the variation of the reflection response of the proposed meta-atom with different crystallization level of GST. It clearly shows the possibility to control the phase of reflected wave over a wide dynamic range of 315$^{\circ}$ while preserving the reflection amplitude above 15\%. By selectively modifying each meta-atom using the proper voltage signal, a broad range of field patterns can be imprinted to the incident light by the metasurface. Here, we are interested in the design and realization of three types of functionalities within a uniquely optimized structure in which: i) the phase-front of light is processed (e.g., for beam focusing) based on geometrical optics concepts and the generalized law of refraction, ii) the amplitude of the wavefront is manipulated (e.g. for switching) through the impedance theory, and iii) polarization state of the beam is controlled (e.g., for polarization conversion) by employing Jones calculations. 
To encode the phase, amplitude, or polarization response associated with each specific functionality, we first discretize the related transverse field distribution (defined by an application specific equation) on the surface of the metasurface and then extract a sufficient sequence of samples. Indeed, each sample represents the reflection value of the corresponding meta-atom with an appropriate GST crystallization level chosen from Figure \ref{fig.eleven crystallization levels}(b). By judiciously arranging those meta-atoms in the lateral dimension, the structure follows field distribution for the desired functionality. It is notable that all functionalities through this study are realized just by adjusting the GST phase through an external stimulus without modifying the structural parameters.

\subsection{Active manipulation of phase distribution}

Here, we focus on the design and implementation of a high numerical aperture (NA) and diffraction-limited metalens with a tunable focal length. 
The required quadratic phase profile follows \cite{yu2011light}
\begin{equation} 
\phi(x)=\dfrac{2\pi}{\lambda}[\sqrt{f^{2}+x^{2}}-f]
\label{lens}
\end{equation}
where $f$ is the focal length, and $\lambda$ is the wavelength in the background medium. In Figure \ref{fig.focusing}(a), the theoretical phase distribution for $f = 20\lambda$ (solid blue line) is compared with the phase profile realized in the full-wave simulation (red circles). Although the discretization is limited to 11 steps (corresponding to the experimentally-demonstrated number of GST crystallization levels \cite{rios2015integrated}), the electric field intensity distribution in Figure \ref{fig.focusing}(b, top) illustrates that focusing occurs at a distance of $20.5\lambda$ in good agreement with the analytical result. Moreover, the normalized field intensity at the focal plane has the full-width at half-maximum (FWHM) of 995 nm and numerical aperture NA = 0.65 justifying the diffraction-limited feature of the designed lens (see Figure \ref{fig.focusing}(c) for comparison with Airy disk). Note that for such a small size lens ($\approx35\lambda$), the efficiency (i.e., the ratio of the reflected power at the focal plane with lateral dimension of three times FWHM to the reflected power from an Au back reflector) is 5.7\%. However, a larger aperture size results in a performance improvement due to the local periodicity effect retention \cite{genevet2017recent}. By manipulating the phase distribution imposed by the metasurface (using an appropriate level of crystallization for each meta-atom), the focal point of metalens can be finely controlled. Figures \ref{fig.focusing}(b, middle) and \ref{fig.focusing}(b, bottom) depict the electric field intensity profiles for metalenses with f = $10\lambda$ and f = $5\lambda$, respectively. In these designs, simulated focal lengths are f = $10.3\lambda$ (with FWHM = 740 nm and NA = 0.86) and f = $5.2\lambda$ (with FWHM = 615 nm and NA = 0.96), respectively. For the sake of clarity, the realized phase profile, the required lateral crystallization distribution, and the calculated depth of focus for the investigated lenses are illustrated in Figure S2 in the Supporting Information. All numerical simulations are in good agreement with the analytical results while negligible discrepancies in the performance primarily originate from the in-plane coupling between the elements of the metasurface array, imprecise discretization of phase profiles, the abrupt phase shift in adjacent meta-atoms, and the finite size of the metalenses. In the Supporting Information, we investigate other optical functionalities, such as phased array antennae, airy beam generator, and beam splitter, that can be realized using the same metasurface structure by engineering the phase-front of light.

\subsection{Dynamic control over the amplitude response}

\begin{figure} [t]
	\centering
	\includegraphics[trim=0cm 6.7cm 0cm 0cm,width=8.3cm,clip]{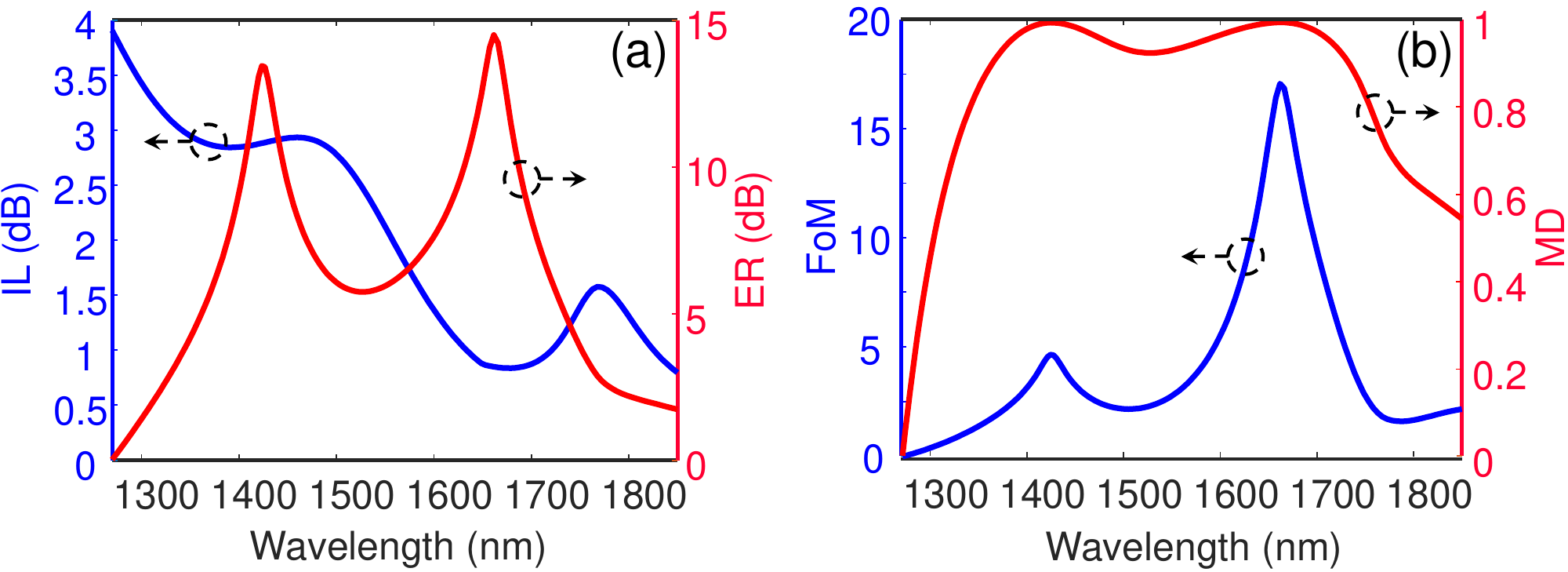}\\
	\caption
	{High-performance behavior of the metasurface in Figure \ref{fig.3D perspective} in optical amplitude modulation. (a) The IL and ER diagrams for the proposed modulator in which ON- and OFF-states correspond to the $L_{c}^{0}$ and $L_{c}^{60}$, respectively. (b) The FoM and MD curves obtained by numerical simulations justify the unprecedented response of the modulator in the telecommunication band.
	}
	\label{fig.modulator}
\end{figure}

By uniformly controlling the crystallization level of GST of each meta-atom, the reflectance spectrum can be independently controlled in a dynamic way. This enables several applications such as switching and perfect absorption. We seek to implement an efficient optical switch with i) low intrinsic insertion loss (IL), which is the device loss in the ON-state, and ii) high extinction ratio (ER), defined as the difference in loss between the ON-state and the OFF-state of the modulator to reduce the bit error rate \cite{yao2013wide,kruger2012design}. However, there is a tradeoff between the IL and the attainable ER, and the ratio of these two parameters is a rational figure-of-merit (FoM) of the modulator. In order to maximize the FoM, we harness the electrically tunable meta-atom working in two distinct crystallization levels of $L_{c}^{0}$ and $L_{c}^{60}$ corresponding to the ON and OFF states, respectively. Figure \ref{fig.modulator} illustrates that the metasurface notably increases the effective FoM and modulation depth (MD) formulated as
\begin{align} 
\textrm{FoM}&=\dfrac{\textrm{ER}}{\textrm{IL}}=\dfrac{-10~\textrm{log}_{10}({|r_\textrm{min}(\lambda)|}/{|r_\textrm{max}(\lambda)|})}{-10~\textrm{log}_{10}({|r_\textrm{max}(\lambda)|})}
,  \nonumber \\ 
\textrm{MD}&=1-\dfrac{{|r_\textrm{min}(\lambda)|^{2}}}{{|r_\textrm{max}(\lambda)|^{2}}}
\label{equ1}
\end{align}
where $r_\textrm{min}(\lambda)$ and $r_\textrm{max}(\lambda)$ are the reflectivity of the modulator in the ON-state and OFF-state at wavelength $\lambda$, respectively. Due to the improved IL and enhanced ER, a significant FoM and MD are obtained over a considerable wavelength range, verifying the high performance of our metasurface-based modulator compared to its counterparts \cite{yao2014electrically, emani2013electrical}. It is noteworthy that high ER, (ER $>$ 7 dB), is preferable for most applications such as large data-rate interconnects and high-sensitivity sensing. However, ER $\sim$ 4 dB can be sufficient for specific applications such as short-distance communication.

\subsection{Tuning the polarization state of incident light}

\begin{figure} [t]
	\centering
	\includegraphics[trim=0cm 0cm 9.5cm 0cm,width=8.3cm,clip]{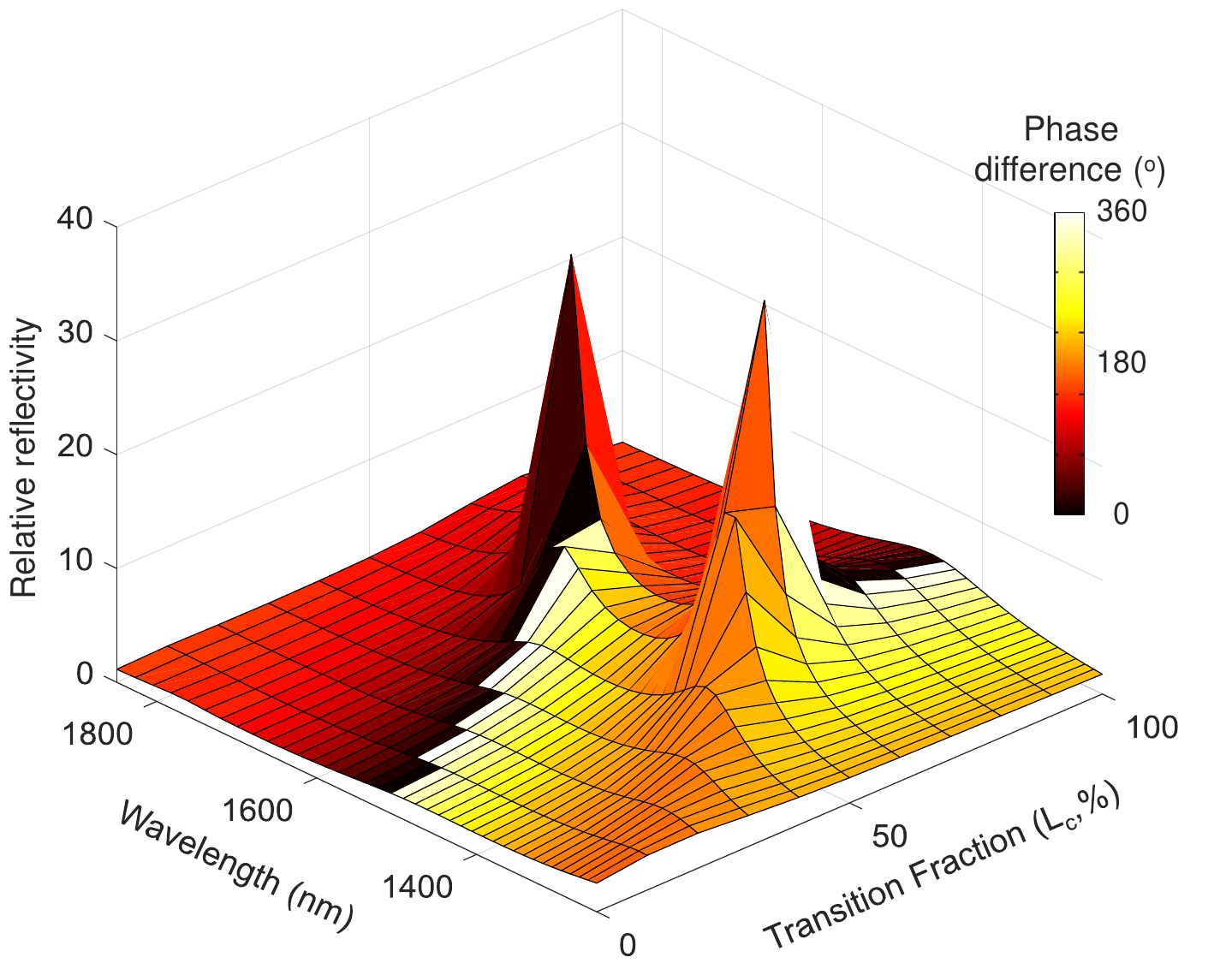}\\
	\caption
	{Polarization conversion using the active metasurface in Figure \ref{fig.3D perspective}. The reflection amplitude ratio ($|r_{x}|/|r_{y}|$) and the relative phase ($\phi_{x}-\phi_{y}$) between the two orthogonally polarized components of the reflected light as a function of the GST crystallization level ($L_{c}^{\%}$) and the wavelength. Different polarization states can be realized in the output by moving on the surface of the plot (i.e., through changing the GST phases).
	}
	\label{fig.polarization}
\end{figure}

By using the relative phase difference between two orthogonal polarization states of the incident beam in Figure \ref{fig.3D perspective}, we realize an active optical waveplate converting an appropriate polarized incident beam to the general form of elliptically polarized light. We consider the Jones matrix representation of the metasurface as
\begin{align}
J_\textrm{metasurface}=
\begin{bmatrix} 
a~\textmd{exp}(i\phi_{x}) & 0 \\
0 & b~\textmd{exp}(i\phi_{y})
\end{bmatrix}
\label{equ9}
\end{align}
where $a$ and $b$ are the reflection amplitudes, and $\phi_{x}$ and $\phi_{y}$ are the reflection phases of the linearly-polarized light parallel and perpendicular to the grating direction in Figure \ref{fig.3D perspective}, respectively. According to Figure \ref{fig.polar plot}, for all GST crystallization levels, $b$ and $\phi_{y}$ can be considered 1 and $\pi$, respectively. As a result, the reflected output field components ($E_\textrm{x,out}$ and $E_\textrm{y,out}$) for a given polarized light can be calculated as 
\begin{align}
\begin{bmatrix} 
E_\textrm{x,out} \\
E_\textrm{y,out}
\end{bmatrix}
&=
\begin{bmatrix} 
a~\textmd{exp}(i\phi_{x}) & 0 \\
0 & \textmd{exp}(i\pi)
\end{bmatrix}
\begin{bmatrix} 
A~\textmd{exp}(i\phi_\textrm{in,x}) \\
\textmd{exp}(i\phi_\textrm{in,y})
\end{bmatrix} \nonumber
\\
&=
\begin{bmatrix} 
aA[\textmd{exp}(i(\phi_{x}+\phi_\textrm{in}))] \\
\textmd{exp}(i(\pi+\phi_\textrm{in,y}))
\end{bmatrix}
\label{equ10}
\end{align}
in which $A$, $\phi_\textrm{in,x}$, and $\phi_\textrm{in,y}$ are the amplitude, x-component, and y-component phases of the incident wave field, respectively. Figure \ref{fig.polarization} represents the relative reflectivity ($|r_{x}|/|r_{y}|$) and the phase difference ($\phi_{x}-\phi_{y}$) between the two orthogonally polarized components of the reflected light in Figure \ref{fig.3D perspective} as a function of the GST crystallization level over a wide spectral range. It is clear that the proposed metasurface generally behaves as a reflective elliptical polarizer just by adjusting the GST crystallization level in the meta-atoms. However, by satisfying particular conditions, several interesting scenarios are expected, e.g.; 1) $\phi_{x}+\phi_\textrm{in}=\pi/2$, $\phi_\textrm{in,y}=\pi$, and $A=a^{-1}$, which results in the left-handed circularly polarized light at the output; 2) $\phi_{x}+\phi_\textrm{in}=3\pi/2$, $\phi_\textrm{in,y}=\pi$, and $A=a^{-1}$ leading to right-handed circularly polarized light at the output; and 3) $\phi_{x}+\phi_\textrm{in}=\pi$ or $2\pi$, and $\phi_\textrm{in,y}=\pi$, which gives rise to a linearly-polarized light at the output. These cases show the strength of our proposed metasurface in engineering the polarization state of light for potential applications in photography, biosensing, and communication applications.

\section{Heat-transfer analysis of the electrically tunable metasurface}

\begin{figure*} [t]
	\centering
	\includegraphics[trim=0cm 2.2cm 7.5cm 0cm,width=13cm,clip]{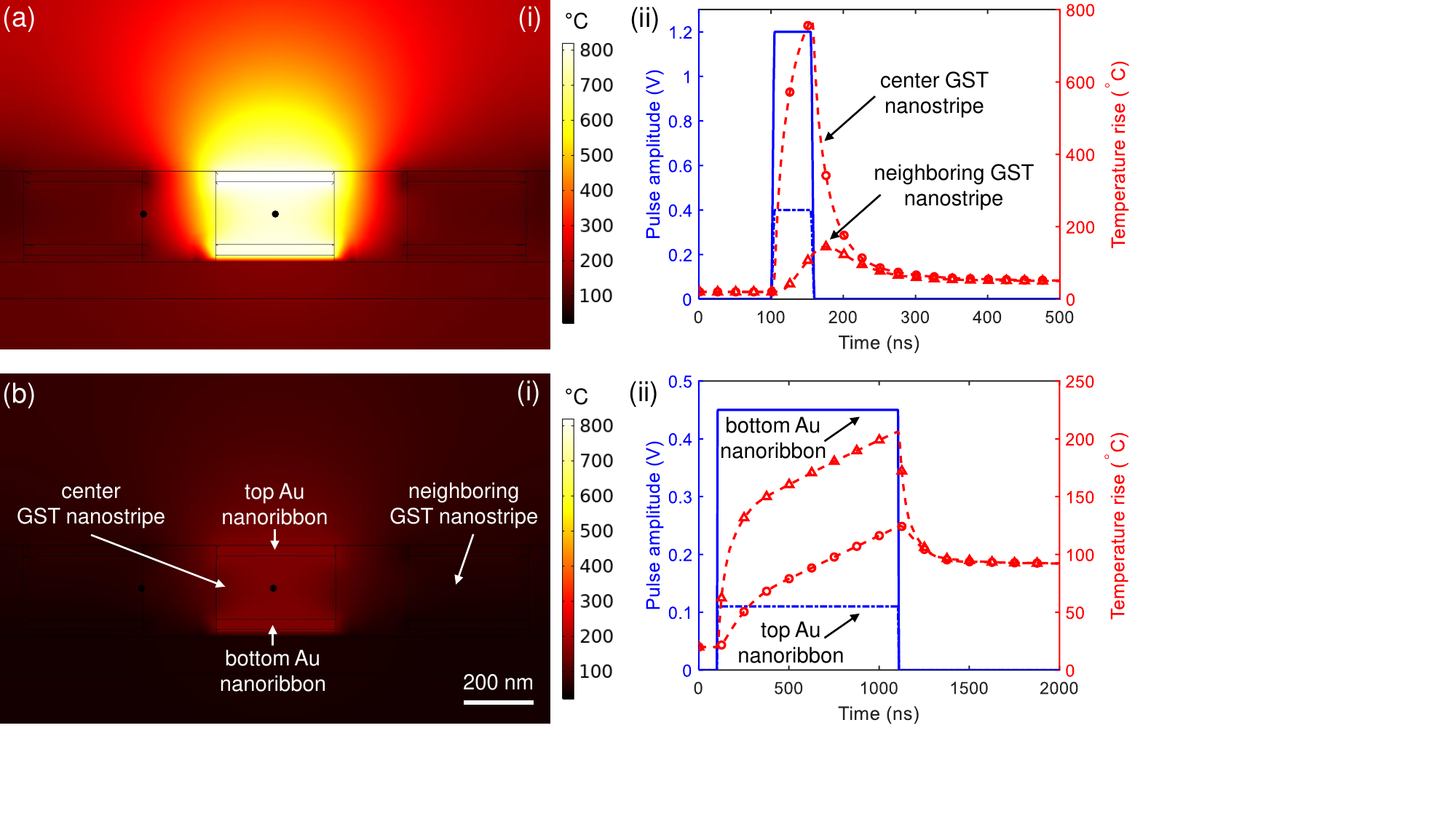}\\
	\caption
	{Simulated electrothermal model for the metasurface with a 1D array of hybrid meta-atoms comprising two stacked gold nanoribbons (as nanoheaters) incorporating a GST nanostripe. The bottom nanoribbon is isolated from the AU back-reflector using a 20 nm-thick SiO$_2$ layer. (a) Simulated temperature distribution at the cross section of the meta-atoms (i.e., black dot) at the end of the (a,i) amorphization pulse (i.e., 150 ns) and (b,i) full crystallization pulse (i.e., 1100 ns). Transient temperature profiles in the center of the middle GST nanostripe as well as on the wall of the neighboring meta-atom upon applying the (a,ii) amorphization and (b,ii) full crystallization pulse. While the whole middle meta-atom in (a) is successfully melted (heated above 630 $^{\circ}$C) followed by a quenching process (with the high cooling rate of $> 10~^{\circ}$C/ns), the maximum temperature experienced by the neighboring meta-atoms is below the glass transition temperature (i.e., $\sim 160 ^{\circ}$C). As shown in (b,ii), the temperature of the neighboring meta-atoms does not go above the glass transition temperature at the end of the crystallization pulse. This ensures addressability at the meta-atom level.
	}
	\label{fig.thermal}
\end{figure*}

Addressing individual subwavelength meta-atoms is an important yet challenging step that can empower dynamic multifunctional metasurfaces to prevail over conventional spatial light modulators (based on liquid crystal or microelectromechanical structures). To figure out the feasibility of local addressing of meta-atoms using Joule heating, heat transfer simulations are carried out for a metasurface with a 1D array of 10-$\mu$m-long meta-atoms comprising two stacked gold nanoribbons (as nanoheaters) incorporating a GST nanostripe. We study the impact of the amorphization pulse on the phase transition of GST by applying 50 ns-long 0.4/1.2 V electrical pulses to the top/bottom nanoheaters. Figure \ref{fig.thermal}a(i) shows the heat profile in a cross section at the center of the array at the end of the pulse. It is evident that the whole area of GST nanostripe can be homogeneously heated above 630 $^{\circ}$C followed by quenching such that GST solidifies in the amorphous state that ensures complete re-amorphization.

To study the impact of the a locally heated meta-atom on the neighboring GST nanostripes located in its proximity, the transitive temperature profiles in the center of the middle GST nanostripe and the side wall of the neighboring cells are depicted in Figure \ref{fig.thermal}a(ii). It shows that during the entire heating process, the temperature of the neighboring GST nanostripes remain well below the onset crystallization temperature (i.e., $\sim$ 160 $^{\circ}$C). Such negligible thermal crosstalk between the neighboring elements guarantees the successful addressability of the phase-change metasurface at the meta-atom level. It is notable that the quenching process with a high cooling rate of $> 1~^{\circ}$C/ns is essential for the amorphization \cite{wuttig2017phase}. This necessitates employing a short electrical pulse for reamorphization. As shown in Figure \ref{fig.thermal}a(ii) this requirement is well satisfied with the proposed material and device platform where the cooling process occurs with an order of magnitude larger rate. For the full crystallization process, we use a low-voltage set pulse (with 1 $\mu$s-long and a peak voltage of 0.11/0.45 V applied to the top/bottom nanoelectrode) that heats up amorphous GST above the crystallization temperature. This pulse is sufficiently long that guarantees full nucleation and formation of monolithic crystalline islands without relying on a stationary DC voltage. By decreasing the voltage amplitude and/or the pulse time duration, arbitrary fractions of amorphous and crystalline states are registered in the material, which results in partial crystalline GST.

\section{Conclusions}

In summary, we theoretically demonstrated the unique capabilities of a single plasmonic-GST metasurface architecture in manipulating the phase, amplitude, or polarization of incident light both spectrally and spatially. We leveraged the two highly confined fundamental plasmonic modes, namely SR-SPP and PR-SPP, coupled to the large refractive index change of GST to effectively control the properties of the reflected light. Accordingly, a diverse set of optical functionalities such as beam focusing, amplitude modulation, and polarization conversion is demonstrated just by adjusting the GST crystallization level in electrically addressable meta-atoms. We showed that the proposed metasurface can relatively overcome the negative effect of amplitude and phase coupling, which limits the multifunctionality of conventional metasurfaces, especially for wideband operation. Such miniaturized, active, and high-speed metasurfaces offer superior platforms for realization of practical metadevices with on-demand control of key properties of light over a large bandwidth for practical applications such as imaging, spectroscopy, holography, and sensing.

\noindent \textbf{Acknowledgement:} 
The work was funded by Office of Naval Research (ONR) (N00014-18-1-2055, Dr. B. Bennet).

\noindent \textbf{Conflict of interest:} 
The authors declare no conflicts of interest regarding this article.



\bibliography{achemso-demo}

\end{document}